\documentclass[12pt,preprint]{aastex}

\usepackage{apjfonts}

\newcommand{\lae}{\mathrel{\raise .4ex\hbox{\rlap{$<$}\lower 1.2ex\hbox{$\sim$}}}}
\newcommand{\gae}{\mathrel{\raise .4ex\hbox{\rlap{$>$}\lower 1.2ex\hbox{$\sim$}}}}

\slugcomment{Submitted to Ap. J.}
\shorttitle{Pictor A Jet Variability}
\shortauthors{Marshall et al.}

\begin{document}

\title{A Flare in the Jet of Pictor A}

\author{H. L. Marshall\altaffilmark{1},
M.J. Hardcastle\altaffilmark{2},
M. Birkinshaw\altaffilmark{3,4},
J. Croston\altaffilmark{9},
D. Evans\altaffilmark{1},
H. Landt\altaffilmark{5},
E. Lenc\altaffilmark{10},
F. Massaro\altaffilmark{3},
E. S. Perlman\altaffilmark{6},
D.A. Schwartz\altaffilmark{3},
A. Siemiginowska\altaffilmark{3},
\L. Stawarz\altaffilmark{7,11},
C. M. Urry\altaffilmark{8},
D.M. Worrall\altaffilmark{3,4},
}
\altaffiltext{1}{Kavli Institute for Astrophysics and Space Research,
Massachusetts Institute of Technology, 77 Massachusetts Ave.,
Cambridge, MA 02139, USA}
\altaffiltext{2}{School of Physics, Astronomy, and Mathematics,
University of Hertfordshire, College Lane, Hatfield, Hertfordshire
UK AL10 9AB}
\altaffiltext{3}{Harvard-Smithsonian Center for Astrophysics,
60 Garden St., Cambridge, MA 02138, USA}
\altaffiltext{4}{Dept. of Physics, University of Bristol, Tyndall Ave., Bristol BS8 1TL, UK}
\altaffiltext{5}{School of Physics, University of Melbourne, Parkville, VIC 3010, Australia}
\altaffiltext{6}{Dept. of Physics and Space Sciences,
Florida Institute of Technology, 150 W. University Blvd., Melbourne, FL, 32901, USA}
\altaffiltext{7}{JAXA/ISAS, 3-1-1 Yoshinodai, Sagamihara, Kanagawa 229-8510, Japan}
\altaffiltext{8}{Department of Physics, Yale University, New Haven, CT 06511, USA }
\altaffiltext{9}{University of Southampton, University Road, Southampton SO17 1BJ, UK}
\altaffiltext{10}{CSIRO Australia Telescope National Facility, P.O. Box 76, Epping, NSW 2121, Australia}
\altaffiltext{11}{Astronomical Observatory, Jagiellonian University, ul. Orla 171, Krak\'ow 30-244, Poland}
\email{hermanm@space.mit.edu,
m.j.hardcastle@herts.ac.uk,
Mark.Birkinshaw@bristol.ac.uk,
J.Croston@soton.ac.uk,
devans@space.mit.edu,
hlandt@unimelb.edu.au,
Emil.Lenc@csiro.au,
fmassaro@head.cfa.harvard.edu,
eperlman@fit.edu,
das@head-cfa.harvard.edu,
aneta@head.cfa.harvard.edu,
stawarz@slac.stanford.edu,
meg.urry@yale.edu,
D.Worrall@bristol.ac.uk
}
\slugcomment{To appear in Ap J Letters}

\begin{abstract}
A {\em Chandra} X-ray imaging observation of the jet in Pictor A showed a feature
that appears to be a flare that faded between 2000 and 2002.
The feature was not detected in a follow-up observation in 2009.
The jet itself is over 150 kpc long and a kpc wide, so finding year-long
variability is surprising.
Assuming a synchrotron origin of the observed high-energy
photons and a minimum energy condition for the outflow, the
synchrotron loss time of the X-ray emitting electrons is of order
1200 yr, which is much longer than the observed variability timescale.
This leads to the possibility that the variable X-ray emission arises
from a very small sub-volume of the jet, characterized by magnetic
field that is substantially larger than the average over the jet.
\end{abstract}


\section{Introduction}

Surveys using the {\em Chandra} X-ray Observatory have been
very successful at detecting knots in quasar jets
\citep[e.g.][]{2005ApJS..156...13M,2004ApJ...608..698S}.
Two mechanisms are generally cited when explaining
the origin of the X-rays: synchrotron emission from high
energy electrons and inverse Compton scattering of
Cosmic Microwave Background photons by low
energy electrons in knots in relativistic bulk motion
along the line of sight \cite[see][for reviews]{2006ARA&A..44..463H,2009A&ARv..17....1W}.
In the synchrotron case, variability might be expected
on a time scale of years due to the short loss times for electrons of sufficient energy
to produce X-rays.  In the inverse Compton model, however, variability time
scales could be much larger, with rise times longer than the
knot light crossing time and decay times dictated by radiative
lifetimes larger than $10^4$ yr.  In both cases, adiabatic
loss time scales should be of order the light crossing time.

Variability studies have generally been limited to nearby objects
such as Cen A \citep{2010ApJ...708..675G} and M 87 \citep{2009ApJ...699..305H}.
In the case of M 87, the HST-1 knot was found to flare by a factor of 50 over a period
of 5 years, becoming much
brighter than the nucleus.
While the M 87 jet is about 7 pc across, the upstream end of the
HST-1 knot is only 1.9 pc across, so a secular increase over 5 yr is reasonable.
However, the increase is not smooth and there is a factor of 2 drop in
less than half a year, so the variation could be up to a factor of 10 faster than
the light travel time across the emission region.
\citet{2003ApJ...586L..41H} explain this by Doppler boosting by a factor
of $\delta \approx 5$.
Both Cen A and M 87 are FR I radio galaxies, which are more common -- and thus
more nearby -- than FR II radio galaxies and quasars with higher power jets.
Knots in high power jets
are larger but less well resolved and variability has not been reported.
For example, one attempt to find variability in the X-ray emission from
a jet in a quasar, 3C 273, yielded a null result
\citep{2006ApJ...648..900J}.

Pictor A is a FR II radio galaxy at a redshift of 0.035.
For $H_0 = 70.5$ km/s/Mpc, 1\arcsec\ corresponds to about 700 pc.
Its pencil-like X-ray jet was found in {\em Chandra} observations
by \citet[][hereafter, WYS01]{2001ApJ...547..740W}, extending 1.9\arcmin\ from the core,
oriented toward the partially resolved western hot spot 4.2\arcmin\ from the core.
They estimated that the jet's typical width was about 2.0\arcsec, or 1.4 kpc.
Based on the assumption that the jet X-ray emission
was produced by the inverse Compton process, and using the radio
galaxy arm-length asymmetry to infer an angle to the line of sight,
they suggested that a plausible scenario was that the magnetic field
in the jet was 2 $\mu$G, about a factor 6 below the equipartition
value, and that the Doppler factor in the jet was $\sim 2.6$. However,
subsequent tighter constraints on the X-ray photon index
\citep{2005MNRAS.363..649H}
mean that the jet would be required to have a
significantly steeper low-energy electron energy index than the lobes
for an inverse Compton model to be viable.
\citet{2005MNRAS.363..649H}
argue that this is implausible, and that the X-ray emission of
the jet must be dominated by the synchrotron process; if so, the X-ray
emission gives us no direct information on the Doppler factor or jet
speed, other than ruling out highly beamed and/or sub-equipartition
models.
While the WYS01 estimate of the angle to the line of sight
remains plausible, there are very large uncertainties associated with
the use of the arm-length asymmetry to estimate $\theta$. As Pictor A
is a broad-line radio galaxy, low-luminosity unified models imply
$\theta \la 45$\arcdeg; the fact that the source is observed as a
lobe-dominated object and the projected linear size of the source
probably requires $\theta \ga 10$\arcdeg.

Examining the archival {\em Chandra} data through 2002
(observation IDs 345, 3090, and 4369), we found
evidence for flares in the jet at $3\sigma$ significance.
The observations were not homogeneous,
with one taken 1\arcmin\ from the core and the other centered at the hotspot
about 4\arcmin\ away, so the point spread function (PSF) at the
core was degraded.
So, to improve the variability test and to check
for new flares, we obtained new images.  Here, we report on the
previous evidence for variability and the results from the new observations.
A detailed analysis of the jet will be the subject of a later paper.

\section{Observations and Data Reduction}

\begin{figure}
  \plotone{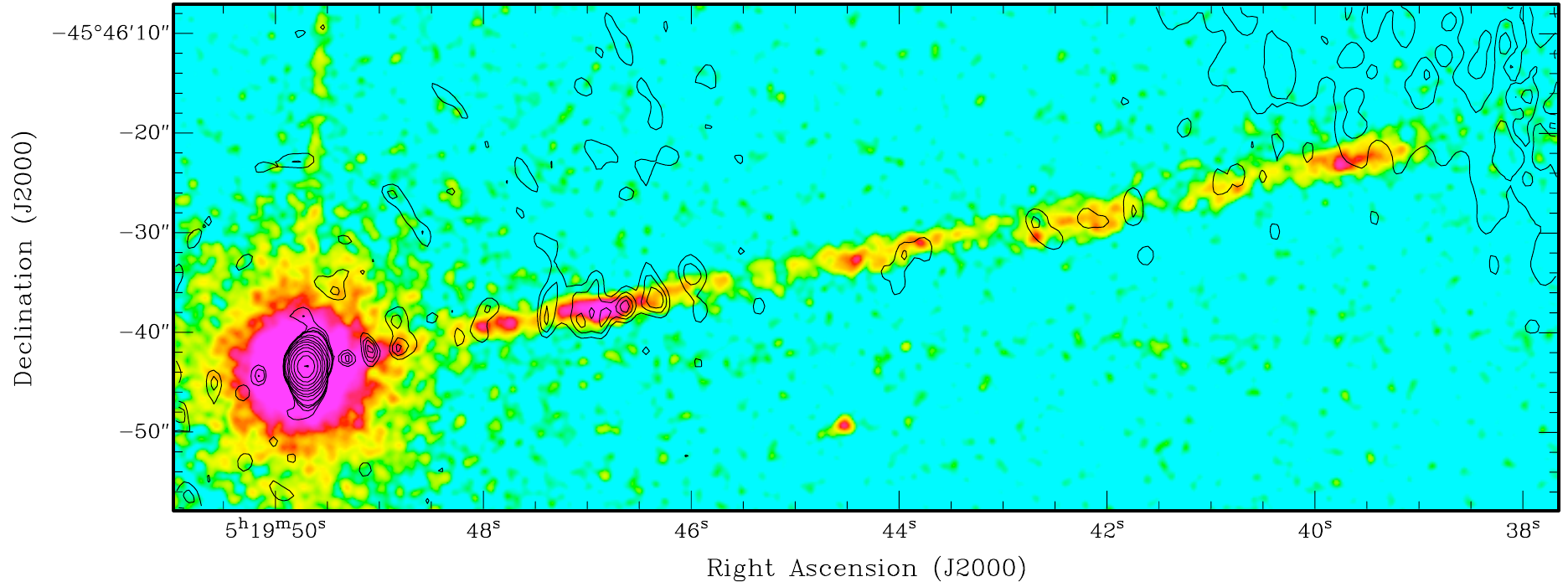}
 \caption{
  X-ray image of the Pictor A jet from all data listed in Table~\ref{tab:observations},
  superposed with radio flux density contours from a new map at 4.8 GHz taken with
  the ATCA.  There are clear associations of radio emission with features
  in the X-ray emission.
  } \label{fig:chandravla}
\end{figure}

Pictor A was observed in 2000, 2002, and 2009 by the {\em Chandra} X-ray
Observatory (see Table~\ref{tab:observations}).
For observation 443, the core was over 6\arcmin\ off-axis, so the core's PSF is
highly extended by comparison to the other observations.  Due to the broadened
PSF and the modest exposure, this observation was not used in the analysis.
Offsets of less than 1\arcmin\ do not degrade the PSF, while at 4.1\arcmin\
off-axis, the PSF is a factor of 2.7 wider than on-axis, based on figures in
the {\em Chandra Proposers' Observatory Guide}.
Unless mentioned otherwise, we selected events from the files processed
using {\tt ciao} in the energy range 0.5-7.0 keV.
A new radio map was obtained using the Australian Telescope Compact Array
(ATCA).  The map, shown in fig.~\ref{fig:chandravla}, shows structure in the inner
jet that was not previously observed, which follows features observed in the
X-ray image.  The radio map will be discussed in more detail in a separate paper.

\begin{figure}[htp]
  \plotone{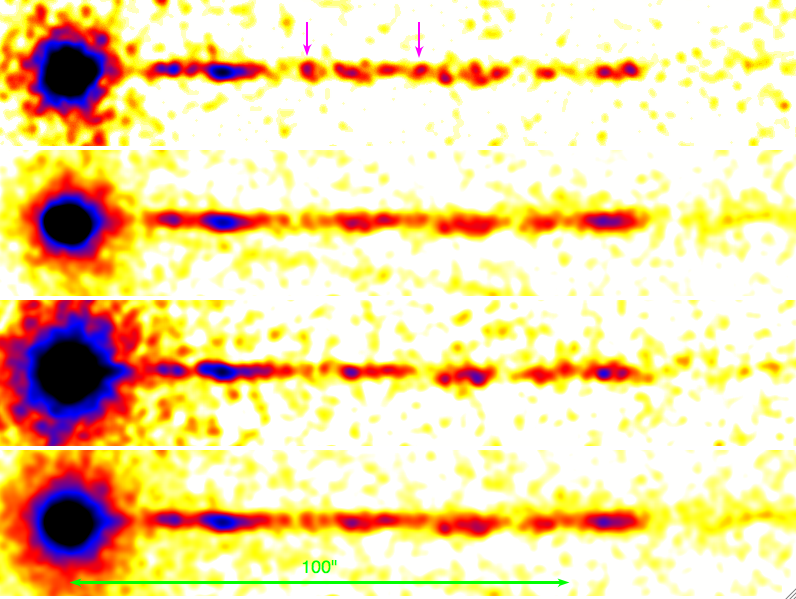}
  \caption{
  X-ray images of the Pictor A jet at several epochs: 2000 (top), 2002 (second from top)
  2009 (second from bottom), and the total (bottom).
  All are rotated by 11.2\arcdeg\ to orient the jet to the right.
  The images were smoothed by a 2D Gaussian with a $\sigma$ of 0.8\arcsec.
  The 2002 image has a readout streak at about 10\arcdeg\ to the jet.
  Possible knot flares are indicated with magenta arrows.
  } \label{fig:images}
\end{figure}

Upon examining the data from 2000 and 2002, we found possible flares in the jet,
as shown in Fig.~\ref{fig:images}.
The most significant feature was at about 48\arcsec\ from the core in
the 2000 image.  A preliminary analysis found 18 counts in obsID 346 from 2000 in
a 3\arcsec$\times$3\arcsec\ region centered on the feature while the combination
of the two observations from 2002 yielded only 15 counts in a longer exposure.
A binomial probability test that there would be $\geqslant$18 of the 33 counts
in the first observation when 7.6 counts were expected (under the null hypothesis
of no variability) gave a probability of significance of $9 \times 10^{-4}$, or an
equivalent Gaussian significance of 3.7$\sigma$.
Accounting for the number of bins examined along the
jet reduced the significance to 2.6$\sigma$.
The flare events were not found to be clustered in time, so ACIS flares
were ruled out.  Similarly, the energy distribution of the events in the flare
was consistent with that of the rest of the jet and significantly different from
that of the background.  So, the events are consistent with originating
in the jet.

The 2009 observations were then combined with the 2002 data for a more
stringent test of the flare at 48\arcsec\ from the core.
We chose a bin size of 1\arcsec\ along the jet to make it straightforward
to find a point source within a lumpy structure (i.e., the jet in its nonvarying ``normal'' state).
A running sum of 3 bins was used to find significant deviations.
In the cross-jet direction, the events were taken from a region within
$\pm 2$\arcsec\  of the centerline set to a position angle of -78.8\arcdeg\
(where positive is E of N).
These values were determined by taking profiles of the jet at
various positions along it and ensuring that a selection this wide
would collect over 90\% of the counts.
The jet ``wiggles'' slightly and broadens along its length
from unresolved to about 2\arcsec\ across, so
the selection region was wider than the PSF across the jet.

\begin{figure}[htp]
  \plotone{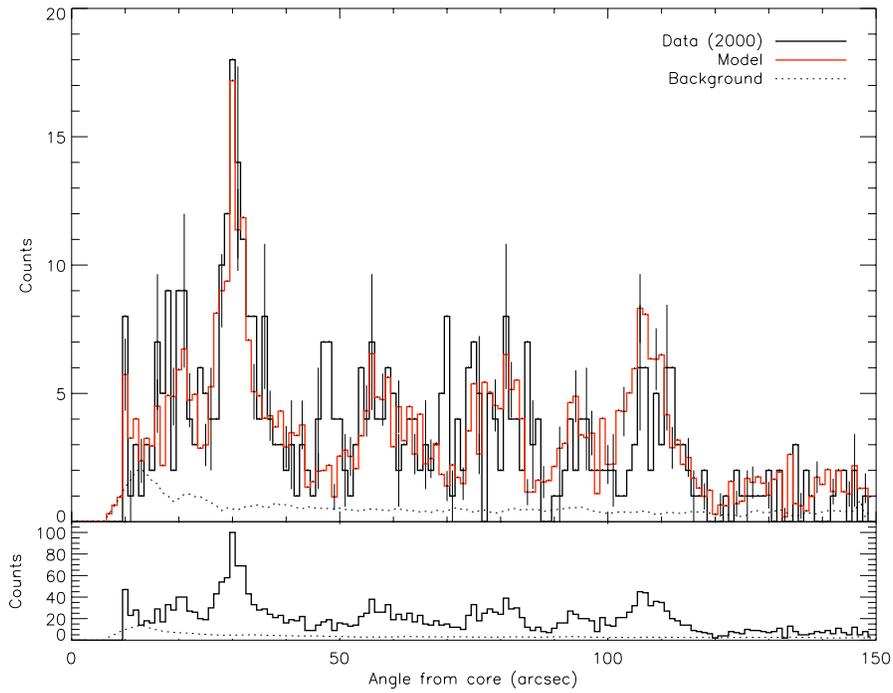}
  \caption{
  {\bf Top}: Count profile of the X-ray emission along the jet for the 2000 observation compared to a model
  based on the remaining observations.  Both the data and the model include smoothed background,
  shown separately.
  The feature at 48\arcsec\ from the core is a possible flare in the jet that was not apparent
  in the 2002 observation.
  Another possible feature is at 70\arcsec\ from the core.
   {\bf Bottom}: Count profile for all observations combined, for reference.  Note that
   the jet is weak but clearly detected beyond 120\arcsec\ from the core.
  } \label{fig:profiletest}
\end{figure}

We compared the profile from the 2000 observation against a
model consisting of two components: a background-subtracted
net model profile and an empirically derived background specific
to the 2000 observation.
The net model profile was constructed from the 2002 and 2009 observations
(see Fig.~\ref{fig:profiletest}).
Backgrounds were determined for each data subset by taking $\pm$2\arcsec\ swaths
at six position angles avoiding chip gaps and other detector features.
The background profiles were subtracted from the count profiles to form the net model,
which was then scaled for comparison to the 2000 profile.
Because the ACIS filter is accumulating a contaminant, the
event energy distribution in 2009 is deficient below 1 keV, making it
inappropriate to normalize those data by a simple factor relating to exposure.
Fortunately, the  event
energy distributions of the inner and outer halves of the 2\arcmin\ long jet are
indistinguishable, indicating that the X-ray spectrum does not vary
significantly along the jet.
So, we chose to normalize the model to match the total counts in
the 2000 profile between 10\arcsec\ and 120\arcsec\ from the core
(after accounting for background), a factor of 0.213.
The background uncertainties end up being negligible but the background
itself is important, contributing as much as the net model does in many locations.

The result is shown in Fig.~\ref{fig:profiletest}.
The most significant deviation between the 2000 data and the
combination of the other data sets is centered at 48\arcsec\ from the core.
The Poisson probability of the deviation was $8.0 \times 10^{-6}$, or an
equivalent Gaussian significance of 4.3$\sigma$.
A similar test between the 2009 and 2002 data yielded no features
more significant than 3.2$\sigma$, so combining these two epochs
is justified {\em a posteriori} and the result provides some
confidence that the analysis method does not overproduce false positive signals.
Accounting for the number of independent trials reduces the significance
to about $3.4\sigma$.  The feature is robust, showing up at 3.4$\sigma$
when testing a running sum of three 0.5\arcsec\ bins; at this binning,
a feature at 70\arcsec\ is significant at 3.3$\sigma$ (before accounting
for the number of trials).  There are only 13.9 net counts in the test
region, so the count rate in the 48\arcsec\ feature
is not well determined: 0.00054 $\pm$ 0.00019 cnt/s.
For a power law spectrum with $\Gamma = 1.94$ and a Galactic column density of
 $5.8 \times 10^{20}$ cm$^{-2}$, as found by WYS01
 for the entire jet, the flux of the flare is
$3.5 \times 10^{-15}$ erg/cm$^2$/s in the 0.5-7.0 keV band.
This flux corresponds to about $2 \times 10^{-15}$ erg/cm$^2$/s
in the 0.5-2.0 keV band, above which there are about 600 X-ray sources
per sq. deg \citep{2001ApJ...562...42T}.  In the 4\arcsec\ wide area of the jet that
was searched for flares from 10 to 120\arcsec\ from the core,
we expect less than 0.02 unrelated X-ray
sources, so the hypothesis that the flare comes from a background
source can be ruled out
at the 98\% confidence level.

It is important to determine if the flare is consistent with a point source.
The jet is resolved in the cross-jet direction
just downstream from the 48\arcsec\ flare location.
Combining all data sets, the cross-jet profile between 50 and 65\arcsec\ from the core
fits a Gaussian with $\sigma = 0.87 \pm 0.05$\arcsec.
An unresolved jet should have a 1D profile comparable to that of the
ACIS readout streak from the unresolved core,
whose profile is a Gaussian with  $\sigma_{\rm psf} = 0.382 \pm 0.015$\arcsec.
Assuming that the true jet profile also matches a Gaussian, then
$\sigma_j = (\sigma^2 - \sigma_{\rm psf}^2)^{1/2}$, giving the jet
FWHM $= 2.35 \sigma_j = 1.83 \pm 0.18$\arcsec.
A proper spatial test would check for a variable point-like feature embedded
in an extended background jet.
For a simple check, we examined the 2000 event list in detail.
We counted the events in concentric 1\arcsec\ and 2\arcsec\ diameter circles, which
should contain 50\% and 90\% of the power.
With 13.9 net counts in the 3\arcsec$\times$4\arcsec\ sample region,
we expect 12.5 counts from the putative point source within the
larger circle and 7 within the smaller one.
We found a location within the sample region
where there are 10 counts within the larger circle
and 4 within the smaller circle, which is
reasonably consistent with the expectation, considering the small number
of counts.
Thus, we conclude that there could have been
a single unresolved knot within the jet that caused the observed flare.

\section{Discussion}

Using the formalism of \cite{2009A&ARv..17....1W}, we estimate the average
equipartition magnetic field in the jet to be $B_{\rm eq} = 17 \mu$G, similar
to the value obtained by WYS01 in the absence of beaming.
Further below we assume that the Pic A jet is indeed close
to the minimum energy condition, as justified by observations
of terminal hotspots and lobes in powerful radio sources
\citep[e.g.][]{2005ApJ...622..797K}.
We measured the jet flux density to be 128 mJy at 1.4 GHz
and assume $\gamma_{\rm min} = 100$, an electron energy index of 2 breaking
to 3 before the X-ray band and
that the radius of the jet is about 1\arcsec, as estimated by WYS01.
For these assumptions, Lorentz factors of the electrons that produce 1 keV X-rays is
$\gamma \approx 7 \times 10^7$.
The corresponding synchrotron loss time, $\tau \approx 1200$ yr, is similar
to the dynamical time scale $t_{\rm dyn} = \ell/c = 2000 $ yr, setting
the knot size to the FWHM of an unresolved source (0.9\arcsec), but both are
much larger than the variability time scale of 2 yr.

Doppler beaming would reduce the estimated intrinsic
value of $B_{\rm eq}$ by a factor of $\delta = 1/[\Gamma (1-\beta \mu)]$,
where $\beta c$ is the speed of the jet moving at angle $\theta = \cos^{-1} \mu$ to the
line of sight, and $\Gamma = (1-\beta^2)^{-1/2}$.
As $\delta$ increases, the efficiency of inverse Compton scattering
of microwave background photons increases rapidly,
affecting the observed X-ray flux. In particular, WYS01 estimated that $\delta
\approx 7$ would allow all the observed X-rays to be produced by the
inverse Compton process with $B=B_{\rm eq}$, although this should be
taken as an upper limit if the jet X-rays are dominated by synchrotron
emission as argued by \citet{2005MNRAS.363..649H}.
Increasing the jet beaming would also affect any
X-ray emission resulting from the synchrotron process.
For a fixed observed photon energy, $E$ (e.g. 1 keV), and the assumed minimum
energy condition, the Lorentz factor of the electrons in the jet rest
frame $\gamma \propto \sqrt{E/B}$ is the same as that inferred
in the observer frame, since the photon energy transforms by
the same factor $\delta$ as $B_{\rm eq}$ scales.
The radiative lifetimes, however, scale as $1/B^2$ and would increase
as $\delta^2$ in the rest frame or $\delta$ in the observed frame,
exacerbating the lifetime discrepancy.

Ignoring beaming for now,
setting $t_{\rm dyn}$ to 2 yr leads to an angular size of the emission
region of $\sim 1$ mas.
For a knot that radiates a fraction $f$ of the total jet flux from a volume
$V \propto \ell^3$, then $B_{\rm eq} \propto (f/\ell^3)^{1/(3+\alpha)}$.
Again fixing the observed photon energy, then $\gamma \propto B^{-1/2}$ and
$\tau \propto B^{-3/2} \propto \ell^{3\xi}/f^\xi$, where $\xi = \frac{3}{6+2\alpha}$.
Setting $\alpha = 0.5$ for our assumed spectral shape gives $\xi = 3/7$.
So, if $\ell$ is $10^3$ smaller, then $\tau$ is a factor of $\sim 7200$ smaller.
For the 48\arcsec\ flare, $f \approx 0.03$, so $\tau \sim 0.7$ yr, which is consistent
with the observed flare.
The local $B$ field would be $\approx 2$ mG, $> 100$ times larger than
the average in the jet.
Thus, in order to explain flares on a timescale of years within a jet
that is about 700 pc in diameter would require
1) X-ray emission from a small knot substantially smaller than the jet cross section
and 2) local magnetic fields substantially larger than the average over the jet.

The energy loss timescale of individual relativistic electrons is
unlikely to be the factor controlling the rate of fading of a flare
since the energy release that is manifest as a flare is also likely to
cause expansion of the emitting region. If we take this emitting
region to be spherical, with radius $R$, and the magnetic field within
it to be tangled, then the optically-thin synchrotron emission from
that region scales as $R^{-4\alpha -2}$ because
expansion decreases both the magnetic field strength and the energies of
the embedded relativistic particles.\footnote{Different
brightness decrease functions arise under different magnetic field
configurations or expansion geometries, but most cases show rapid loss
of power under adiabatic losses, at sufficiently high observing
frequency.}  If the flaring region had
$\alpha = 0.5$, then an expansion by 50\% would cause it to fade 
by a factor 5 and become undetectable against the larger-scale jet
emission. Such an expansion is feasible if the flaring region is only
2~ly in diameter.

As VLBI observers expand their maps to include knots at one
arcsec scales, some knots are showing mas-scale structure.
For example, \cite{2009ApJ...695..707G} found a
very compact hot spot 40 kpc from the core of a quasar and \cite{2010arXiv1002.2588C}
detected the HST-1 knot in VLBI observations.
Examining more cases might well prove to be a fruitful endeavor.
Imaging at even 0.1\arcsec\ scales with the {\em Hubble} Space
Telescope and future generations of X-ray telescopes could well
show that jet knots have substantial substructure.

\acknowledgments

Support for this work was provided in part by the National Aeronautics and
Space Administration (NASA) through the Smithsonian Astrophysical Observatory (SAO)
contract SV3-73016 to MIT for support of the Chandra X-Ray Center (CXC),
which is operated by SAO for and on behalf of NASA under contract NAS8-03060.
Support was also provided by NASA under contract NAS 8-39073 to SAO.
This research has made use of the NASA/IPAC Extragalactic
Database (NED) which is operated by the Jet Propulsion Laboratory,
California Institute of Technology, under contract with the
National Aeronautics and Space Administration.
\L . S. is grateful for the support from Polish MNiSW through the grant
N-N203-380336.

{\it Facilities:} \facility{CXO(ACIS)}, \facility{ATCA ()}

\clearpage

\clearpage

\begin{deluxetable}{lccc}
 \tablecolumns{4}
 \tablewidth{0pc}
 \tablecaption{{\em Chandra} Observations of Pictor A \label{tab:observations} }
 \tablehead{
\colhead{{\em Chandra}} & \colhead{Off-axis Angle} & \colhead{Live Time} 
	& \colhead{Date} \\
\colhead{Obs ID} & \colhead{(\arcmin)} & \colhead{(s)} 
	& \colhead{}}
\startdata
   346 &   0.7 &  25733 & 2000-01-18 \\
   443 &   6.4 &   5058 & 2000-06-28 \\
  3090 &   4.1 &  46362 & 2002-09-17 \\
  4369 &   4.1 &  49123 & 2002-09-22 \\
 11586 &   0.8 &  14257 & 2009-12-12 \\
 12039 &   0.8 &  23737 & 2009-12-07 \\
 12040 &   0.8 &  17319 & 2009-12-09 \\
\enddata
\end{deluxetable}

\end{document}